\documentclass[showpacs,aps,twocolumn,prl]{revtex4}
\usepackage[dvips]{graphicx}
\usepackage{bm}% bold math

\begin{document}

\title{
{\em Ab initio} Study of Valley Line on a Total-Energy Surface\\
for Zone-Center Distortions of Ferroelectric Perovskite Oxides BaTiO$_{3}$ and PbTiO$_{3}$
}

\author{Takeshi~Nishimatsu$^1$}
\author{Takatoshi~Hashimoto$^{1,2}$}
\author{Hiroshi~Mizuseki$^1$}
\author{Yoshiyuki~Kawazoe$^1$}
\author{Atsushi~Sasaki$^2$}
\author{Yoshiaki~Ikeda$^2$}

\affiliation{$^{1}$Institute for Materials Research (IMR),
  Tohoku University, Sendai 980-8577, Japan\\
  $^{2}$NEC TOKIN Corporation, Sendai 982-8510, Japan}

\begin{abstract}
An {\it ab initio} structure optimization technique is newly developed
to determine the valley line on a total-energy surface for
zone-center distortions of
ferroelectric perovskite oxides
and is applied to
barium titanate (BaTiO$_{3}$) and
lead titanate (PbTiO$_{3}$).
The proposed technique is an improvement over
King-Smith and Vanderbilt's scheme [Phys. Rev. B {\bf 49}, 5828 (1994)]
of evaluating total energy as
a function of the amplitude of atomic displacements.
The results of numerical calculations show that
total energy can be expressed as a fourth-order function
of the amplitude of atomic displacements in BaTiO$_3$
but not in PbTiO$_3$.
\end{abstract}

\date{\today}

\pacs{
% In 77. DIELECTRICS, PIEZOELECTRICS, AND FERROELECTRICS AND THEIR PROPERTIES:
77.84.-s,   % Dielectric, piezoelectric, ferroelectric, and antiferroelectric materials
77.80.Bh,   % Phase transitions and Curie point
% In 63. Lattice dynamics
63.70.+h,   % Statistical mechanics of lattice vibrations and displacive phase transitions
% In 02. Mathematical methods in physics
02.60.Pn    % Numerical optimization
}

\maketitle

%%%%%%%%%%%%%%%%%%%%%%%%%%%%%%%%%%%%%%%%%%%%%%%%%%%%%%%%%%%%%%%%%%%%%%%%%%%%%%%%

\begin{figure}
  \includegraphics[width=50mm]{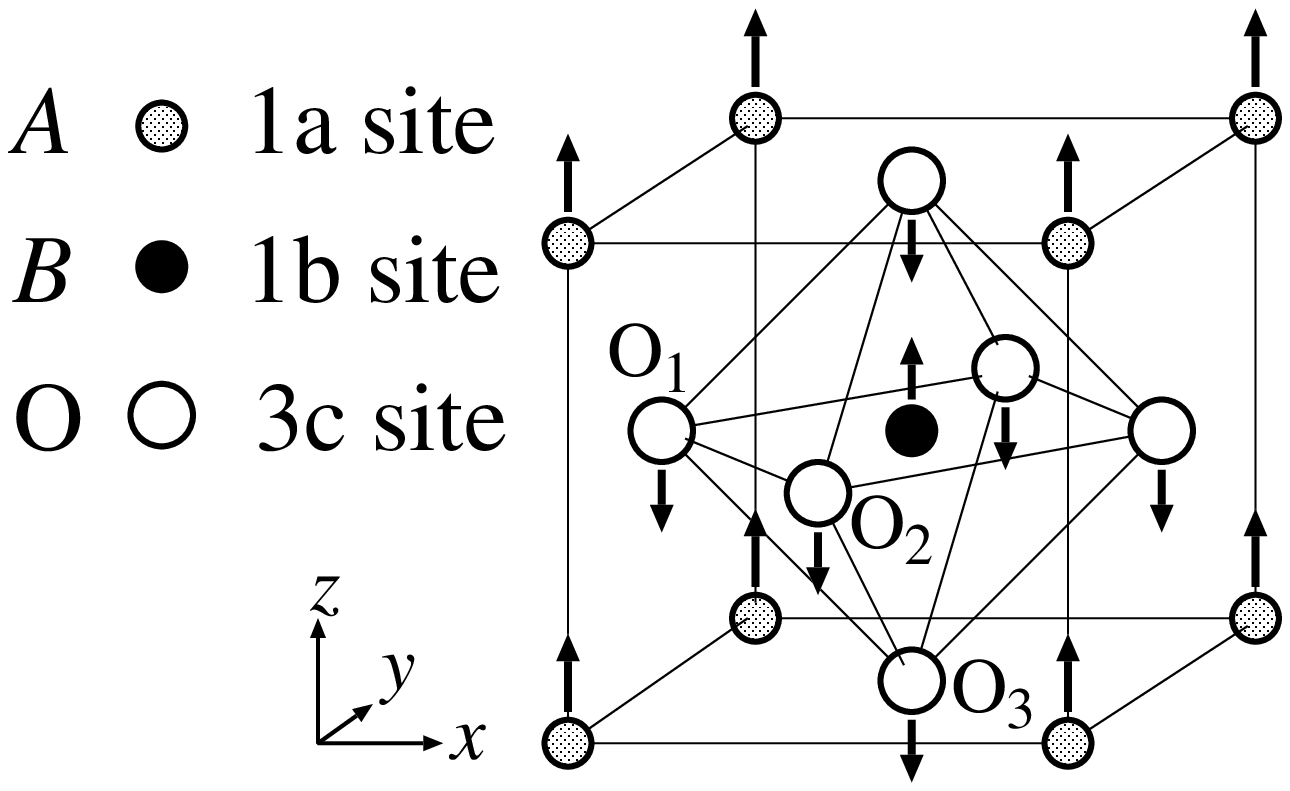}
  \caption{Cubic crystal structure of perovskite oxide $AB$O$_3$.
    The cubic-tetragonal displacive transition accompanies
    atomic displacements $v_z^{\tau}$
    ($\tau$=$A$, $B$, O$_1$, O$_2$, O$_3$) indicated by arrows.}
  \label{fig:structure}
\end{figure}

The $AB$O$_3$ perovskite oxides,
such as barium titanate (BaTiO$_3$) and lead titanate (PbTiO$_3$),
are extremely important ferroelectric materials.
Because of their piezoelectricity below
Curie temperatures (408 and 763~K, respectively~\cite{Subbarao:1973})
and large dielectric constants,
they are widely used in various applications
and they have been extensively studied
experimentally and theoretically.
At room temperature,
BaTiO$_3$ and PbTiO$_3$ have tetragonal ($C_{4v}^1$) structures
and spontaneous polarization.
The tetragonal structure
can be understood to be a result of displacive transitions
from their high-temperature cubic ($O_h^1$) structures,
as depicted in Fig.~\ref{fig:structure}.
The spontaneous polarization is a result of
the displacements of cations $A$ and $B$ and three O$^{2-}$'s
to the opposite directions.
While these two classic examples of ferroelectric perovskite oxides,
BaTiO$_3$ and PbTiO$_3$, have similar properties,
such as crystal structure and
unit-cell volume (64.4 and 63.4~\AA$^3$, respectively),
they have significantly different structural parameters and properties,
such as the ratio of lattice constants $c/a$ (1.01 and 1.06, respectively),
atomic displacements
(in BaTiO$_3$ the atomic displacement of $B\!=$Ti
is larger than that of $A\!=$Ba,
but in PbTiO$_3$ that of $A\!=$Pb is larger than that of $B\!=$Ti),
dielectric constants, and piezoelectric constants.
Cohen used the all-electron, full-potential,
linearized augmented plane-wave (FLAPW) method
to study the atomic and electronic structures
of BaTiO$_3$ and PbTiO$_3$~\cite{Cohen:Nature:1992}.
He pointed out that
the hybridization between Ti 3d and O 2p
is strong and essential
to weaken the short-range repulsions
and allow the ferroelectric transition
in both BaTiO$_3$ and PbTiO$_3$.
He also pointed out that
the hybridization between Pb 6s and O 2p
is strong so that it modifies
the ground state and nature of the transition of PbTiO$_3$ by
(1) increasing the ferroelastic strain that couples with
the ferroelectric distortions or
(2) hybridizing with the valence states,
leading indirectly to changes in the Ti-O interactions,
whereas the interaction between Ba and O is
almost ionic in tetragonal BaTiO$_3$.
Recently,
using synchrotron orbital radiation high-energy X-rays,
Kuroiwa {\it et al.} confirmed
the covalent bonding
between Pb-O and
the ionic bonding
between Ba-O~\cite{Kuroiwa:X-ray:Pb:O:Covalency:2001}.

\begin{figure}
  \includegraphics[width=50mm]{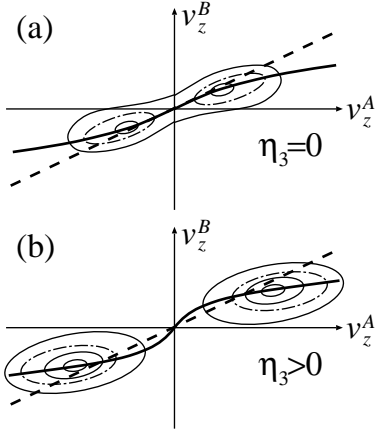}
  \caption{The total-energy surface for
    zone-center distortions of
    ferroelectric perovskite oxides
    are schematically illustrated with contour lines
    for two strains (a)~$\eta_3 = 0$ and (b)~$\eta_3 > 0$
    in the two-dimensional subspace $(v_z^A, v_z^B)$ of
    the atomic-displacement space
    $(v_z^A, v_z^B, v_z^{\rm O_1}, v_z^{\rm O_2}, v_z^{\rm O_3})$.
    Thick solid lines are the valley lines for fixed $\eta_3$'s.
    Dashed lines show the {\it direction} of
    the $\Gamma_{15}$ soft-mode eigenvector at zero strain.
    Note that the {\it direction} is tangential to
    the valley line at $v_z^\tau=0$ for $\eta_3=0$,
    but this is not the case for $\eta_3\neq 0$.}
  \label{fig:ValleyLine}
\end{figure}

To investigate ferroelectric perovskite oxides,
it is useful to estimate the total-energy surface
for zone-center distortions accurately
under absolute zero temperature by {\it ab initio} calculations.
King-Smith and Vanderbilt studied the total-energy surface
using ultrasoft-pseudopotentials and
a plane-wave basis set~\cite{King-Smith:V:1994}.
From the full symmetric cubic perovskite structure,
they gave the displacements $v_{\alpha}^{\tau}$
of atoms $\tau$ (=$A$, $B$, O$_1$, O$_2$, O$_3$)
in the Cartesian directions of $\alpha~(=x,y,z)$
along the $\Gamma_{15}$ soft-mode normalized
eigenvectors $\bm{\xi}_\alpha$ as
\begin{equation}
  \label{eq:Eigenvector}
{\bf v}_\alpha=
\left(
  \begin{array}{c}
    v_\alpha^A         \\
    v_\alpha^B         \\
    v_\alpha^{\rm O_1} \\
    v_\alpha^{\rm O_2} \\
    v_\alpha^{\rm O_3}
  \end{array}
\right)
= u_\alpha\bm{\xi}_\alpha
= u_\alpha
  \left(
    \begin{array}{c}
      \xi_\alpha^A         \\
      \xi_\alpha^B         \\
      \xi_\alpha^{\rm O_1} \\
      \xi_\alpha^{\rm O_2} \\
      \xi_\alpha^{\rm O_3}   
    \end{array}
  \right)\ ,
\end{equation}
with the scalar soft-mode amplitude $u_\alpha$.
Under the condition that the strain components
$\eta_i$~($i=1,\ \dots,\ 6$; Voigt notation)
minimize the total energy for each $u_\alpha$,
they expressed the total-energy surface as
\begin{equation}
  \label{eq:King-Smith-and-Vanderbilt}
  E_{\rm tot}
  = E^{0}
  + \kappa u^2
  + \alpha ' u^4
  + \gamma ' (u_x^2 u_y^2 +
              u_y^2 u_z^2 +
              u_z^2 u_x^2)\ ,
\end{equation}
where $u^2 = u_x^2 + u_y^2 + u_z^2$,
$E^{0}$ is the total energy for the cubic structure,
$\kappa$ is one-half of the eigenvalue of the soft mode,
and $\alpha '$ and $\gamma '$ are the constants determined from
coupling constants between
atomic displacements and strains.
Although their expression veritably describes essential
properties of atomic displacements coupling with strains,
it may misestimate (a) the energy gain of the lattice distortion
from the cubic structure
to the tetragonal equilibrium structure due to a deviation
of atomic displacements
from the soft-mode $\bm{\xi}_\alpha$ {\em direction}
as illustrated in Fig.~\ref{fig:ValleyLine},
as well as (b) the total energy particularly
when the amplitude of atomic displacements
$u_z$ is larger than the equilibrium value.
Also, (c) $\alpha '$ and $\gamma '$ in eq.~(\ref{eq:King-Smith-and-Vanderbilt})
strongly depend on coupling constants between
atomic displacements and strains,
but the coupling constants,
$B_{1xx}=\partial^3 E/\partial\eta_1\partial u_x^2$ for instance,
are {\em not} always constant through out
all the way from the cubic structure to the distorted equilibrium structure,
and thus they are difficult to determine.
These issues, (a)-(c), arise
because King-Smith and Vanderbilt's expression is restricted
within the atomic displacements corresponding to
the $\Gamma_{15}$ soft-mode eigenvector at zero strain
and also within the fourth-order function of $u_\alpha$,
as in eq.~(\ref{eq:King-Smith-and-Vanderbilt}).
In this work, therefore, we redefine the amplitude of atomic displacements as
\begin{equation}
  \label{eq:redefinition}
     u_{\alpha}=\sqrt{
                 \big( v_\alpha^A         \big) ^2
                +\big( v_\alpha^B         \big) ^2
                +\big( v_\alpha^{\rm O_1} \big) ^2
                +\big( v_\alpha^{\rm O_2} \big) ^2
                +\big( v_\alpha^{\rm O_3} \big) ^2}\ ,
\end{equation}
and evaluate the total energy as a function of $u_{\alpha}$
under the condition that
$v_{\alpha}^{\tau}$
and
$\eta_i$
minimize the total energy for each $u_\alpha$
using an {\it ab initio} norm-conserving pseudopotential method
and geometric optimization.
Note that
the translational displacement should be omitted both
in eq.~(\ref{eq:Eigenvector}) and
in eq.~(\ref{eq:redefinition}),
i.e., the condition that
\begin{equation}
  \label{eq:zero}
  v_\alpha^A         +
  v_\alpha^B         +
  v_\alpha^{\rm O_1} +
  v_\alpha^{\rm O_2} +
  v_\alpha^{\rm O_3} = 0
\end{equation}
should be satisfied.
Although, below room temperature, BaTiO$_3$ exhibits
tetragonal--orthorhombic~($C_{2v}^{14}$)--rhombohedral~($C_{3v}^5$)
structure transitions
and the total energy can be defined as a function of $u_x, u_y$, and $u_z$,
we restrict our argument, in this paper,
only to the cubic--tetragonal distortion;
$E_{\rm tot} \equiv E_{\rm tot}(u_x=0, u_y=0, u_z)$
and
we optimize the strains $\eta_1=\eta_2$ and $\eta_3$
while $\eta_4$, $\eta_5$, and $\eta_6$ are kept at zero.

Minimization of total energy
under the constant-$u_{\alpha}$ constraint,
i.e., on the constant-$u_{\alpha}$ sphere,
is carried out iteratively.
Suppose that at the previous iteration indexed $k$
we calculated a gradient ${\bf g}$,
i.e., forces exerted on atoms,
and a symmetric Hessian matrix $B$,
i.e., the second derivative of total energy
$B_{\sigma\tau}=d^2E_{\rm tot}/dv_z^\sigma dv_z^\tau$,
at displacement vector ${\bf v}_{z,k}$.
In the neighborhood of  ${\bf v}_{z,k}$,
total energy has the form
\begin{eqnarray}
  \label{eq:EtotAroundVzk}
  \nonumber
  E_{\rm tot}({\bf v}_{z})=
  \frac{1}{2}
   {}^t({\bf v}_{z}-{\bf v}_{z,k})B({\bf v}_{z}-{\bf v}_{z,k})\\*
  +{}^t({\bf v}_{z}-{\bf v}_{z,k}){\bf g}
  +E_{\rm tot}({\bf v}_{z,k})\ ,
\end{eqnarray}
where $t$'s indicate the transpose of vectors.
We first minimize eq.~(\ref{eq:EtotAroundVzk}) on a plane that
includes ${\bf v}_{z,k}$ and is perpendicular to
${\bf n}={\bf v}_{z,k}/|{\bf v}_{z,k}|$,
i.e., on a plane that touches the constant-$u_{\alpha}$ sphere
at ${\bf v}_{z,k}$.
This constraint can be expressed as
\begin{equation}
  \label{eq:Plane}
  {}^t({\bf v}_{z}-{\bf v}_{z,k}){\bf n}=0\ .
\end{equation}
The method of Lagrange multipliers specifies that the gradients of
eq.~(\ref{eq:EtotAroundVzk}) and eq.~(\ref{eq:Plane}) must be proportional.
Thus, a trial displacement ${\bf v}'_{z,k+1}$
for the next iteration $k+1$ can be determined
by solving
\begin{equation}
  \label{eq:MustBePropotional}
  B({\bf v}_{z}-{\bf v}_{z,k})+{\bf g}=\lambda{\bf n}\ ,
\end{equation}
where $\lambda$ is a Lagrange multiplier.
We can determine the Lagrange multiplier
$\lambda={}^t{\bf n}B^{-1}{\bf g}/
         {}^t{\bf n}B^{-1}{\bf n}$,
and now we obtain
\begin{equation}
  \label{eq:SlightlyLarge}
  {\bf v}'_{z,k+1}
  ={\bf v}_{z,k}
  -B^{-1}({\bf g}-\frac{{}^t{\bf n}B^{-1}{\bf g}}
                              {{}^t{\bf n}B^{-1}{\bf n}}
                         {\bf n}
         )\ ,
\end{equation}
but $|{\bf v}'_{z,k+1}|$ may be slightly different from $u_z$.
Hence, we use
\begin{equation}
  \label{eq:Final}
  {\bf v}_{z,k+1}=\frac{{\bf v}'_{z,k+1}}{|{\bf v}'_{z,k+1}|}u_z
\end{equation}
as a trial displacement for the next iteration.
The strains $\eta_1=\eta_2$ and $\eta_3$, i.e.,
lattice constants $a$ and $c$,
are optimized simultaneously in the iterations.
We continue this iterative scheme of
optimizing the atomic and lattice structure
until the differences in total energies become less than
$10^{-7}$ Hartree twice successively.

In the case of chemical reactions of molecules,
a reaction path can be defined as the steepest-descent path from
a transition structure
down to reactants and
down to products, i.e., the valley line~\cite{Fukui:1981}.
Although there are some techniques
to determine the chemical reaction path~\cite{BernhardSchlegel:1995},
we expect
our optimization technique
to be accurate enough and
to give a simple approximation
for the $AB$O$_3$ perovskite oxides.

We use the ABINIT package~\cite{Gonze:ABINIT.ComputMaterSci:2002}
for all {\it ab initio} calculations.
Bloch wave functions of electrons are expanded into
plane waves with a cut-off energy of 60~Hartree,
using Teter's extended norm-conserving pseudopotentials~\cite{Teter:Pseudopotential:1993}.
The pseudopotentials include
O 2s and 2p,
Ti 3s, 3p, 3d and 4s,
Ba 5s, 5p and 6s,
and
Pb 5d, 6s and 6p
as valence electrons.
Bloch wave functions are sampled on an
$8\!\times\! 8\!\times\! 8$ grid of $k$-points
in the first Brillouin zone.
The grid is reduced to 40 irreducible
$k$-points under $C_{4v}^1$ symmetry.
The exchange-correlation energy is treated
within the local density approximation (LDA).
As the parametrized correlation energy,
we use Teter's rational polynomial parametrization~\cite{Goedecker:T:H:1996},
which reproduces the results
obtained by Ceperley and Alder~\cite{CeperleyAlder}.
The electronic states are calculated by the iterative scheme
to reach a tolerance of convergence that requires
differences of forces to be less
than $5\!\times\! 10^{-7}$ Hartree/Bohr twice successively.

The presently calculated ratios of lattice constants $c/a$ and
the amplitudes of atomic displacements $u_z$
for the tetragonal equilibrium structures are
$c/a=$ 1.0029 and $u_{z,{\rm eq}}=$ 0.0975~Bohr for BaTiO$_3$ and
$c/a=$ 1.0284 and $u_{z,{\rm eq}}=$ 0.6087~Bohr for PbTiO$_3$.
The experimentally observed values are
$c/a=$ 1.0086 and $u_{z,{\rm eq}}=$ 0.2638~Bohr for BaTiO$_3$~\cite{Kwei:L:B:C:1993} and
$c/a=$ 1.0649 and $u_{z,{\rm eq}}=$ 0.9076~Bohr for PbTiO$_3$~\cite{Kuroiwa:X-ray:Pb:O:Covalency:2001}.
Thus, calculations underestimate the experimental values.
Calculated energies gained through cubic to tetragonal distortions
are 0.386~meV in BaTiO$_3$ and 37.5~meV in PbTiO$_3$.
Although energy gain and experimentally observed transition temperature
(408~K $=$ 35.2~meV for BaTiO$_3$ and 763~K $=$ 65.8~meV for PbTiO$_3$)
cannot be compared directly,
the energy gains also seem to be underestimated.
These underestimations are because
the large sensitivity to volume makes
the volume errors in LDA
unusually important~\cite{Cohen:Nature:1992}.
Nevertheless, we believe that calculations using LDA
clarify some trends of displacive transitions of perovskite oxides.

\begin{figure}
  \includegraphics[width=60mm]{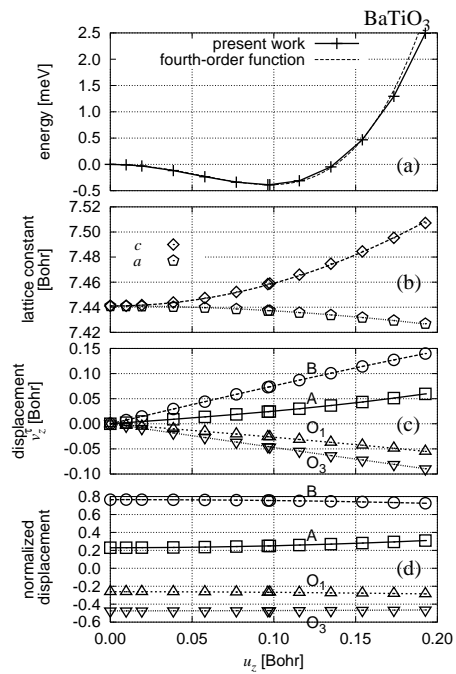}
  \caption{(a)~Calculated total energy in meV as a function of
    $u_z$ in Bohr for BaTiO$_3$ ($+$'s connected with solid lines)
    compared with
    a fourth-order function that passes through
    the calculated energy minimum at $u_z=0.0975$ Bohr and the maximum at $u_z=0$ (dashed line).
    Zero of the energy scale is placed at
    the total energy of the cubic structure when $u_z=0$.
    (b)~Lattice constants $a$ and $c$ in Bohr as functions of $u_z$
    fitted by quadratic functions
    drawn with a dotted line and a dashed line, respectively.
    (c)~Atomic displacements $v_z^A, v_z^B, v_z^{\rm O_1}=v_z^{\rm O_2},
    v_z^{\rm O_3}$ as functions of $u_z$.
    (d)~Normalized atomic displacements $v_z^\tau/u_z$ as functions of $u_z$.
    The $\Gamma_{15}$ soft-mode eigenvector $\bm{\xi}_z$
    calculated by the frozen phonon method
    is additionally shown at $u_z=0$.}
  \label{fig:TotalEnergyFunctionBaTiO3}
\end{figure}

Figure~\ref{fig:TotalEnergyFunctionBaTiO3} shows
the obtained results for BaTiO$_3$.
We can see that total energy is well
expressed as a fourth-order function
of the amplitude of atomic displacements $u_z$
(Fig.~\ref{fig:TotalEnergyFunctionBaTiO3}(a)).
Lattice constants $a$ and $c$ are
also well fitted with the quadratic function of $u_z$
(Fig.~\ref{fig:TotalEnergyFunctionBaTiO3}(b)).
These results suggest that
for the compound with the small spontaneous distortion
the {\it direction} of atomic displacements
remains almost equal to the soft-mode eigenvector $\bm{\xi}_z$
through out the speculating range of atomic displacements,
as shown in Fig.~\ref{fig:TotalEnergyFunctionBaTiO3}(d),
and thus the total energy can be well expressed
within the fourth order expansion.

\begin{figure}
  \includegraphics[width=60mm]{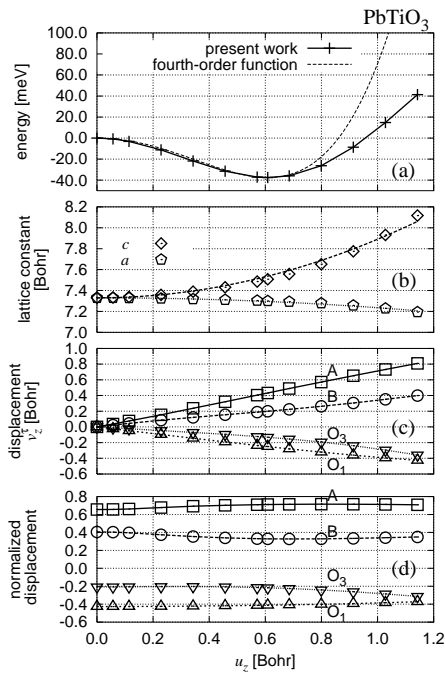}
  \caption{(a)-(d) Same as Fig.~\ref{fig:TotalEnergyFunctionBaTiO3}
    except the energy minimum is at $u_z=0.6087$ for PbTiO$_3$.
    Note that scale of energy and displacements are larger than
    that in Fig.~\ref{fig:TotalEnergyFunctionBaTiO3} and
    that the order of atomic displacements of PbTiO$_3$ in (c) and (d)
    is $A$, $B$, O$_3$, and O$_1$,
    from top to bottom, which is different from that of BaTiO$_3$.}
  \label{fig:TotalEnergyFunctionPbTiO3}
\end{figure}

In PbTiO$_3$, on the contrary,
total energy cannot be
expressed as a fourth-order function of $u_z$,
as shown in Fig.~\ref{fig:TotalEnergyFunctionPbTiO3}(a).
We can see in Fig.~\ref{fig:TotalEnergyFunctionPbTiO3}(b)
that lattice constant $c$ is not well fitted
by the quadratic function of $u_z$.
These are due to the deviation of atomic displacements
from the soft-mode $\bm{\xi}_\alpha$ {\em direction}
as $u_z$ becomes larger (Fig.~\ref{fig:TotalEnergyFunctionPbTiO3}(d));
from $u_z=0$ to around $u_z=0.6087$,
at which the total energy becomes minimum,
the normalized displacement of $A\!=$Pb increases
and that of $B\!=$Ti decreases.
Then, beyond $u_z=0.6087$,
that of O$_3$ decreases while that of O$_1$ slightly increases.
This result is the first clarification
of the non-fourth-order behavior of the total-energy surface,
though it was predicted
for the compounds with the largest spontaneous distortion~\cite{King-Smith:V:1994},
but has been difficult to formalize
with coupling between atomic displacements and strains.

\begin{figure}
  \includegraphics[width=60mm]{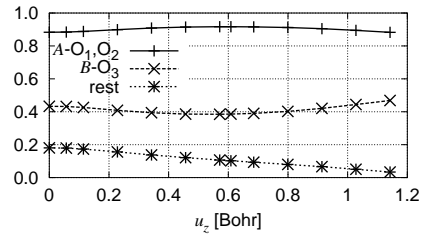}
  \caption{Normalized atomic displacements
    of PbTiO$_3$ are decomposed into three modes:
    the $A$-O$_1$,O$_2$ closing mode (solid line),
    the $B$-O$_3$ closing mode (dashed line),
    and the rest mode (dotted line)
    as functions of $u_z$ in Bohr.}
  \label{fig:PbTiO3Mode}
\end{figure}

The conventional Slater mode~\cite{Slater:1950}
$\bm{\xi}_{\rm Slater} = {}^t$(0 3 $-1$ $-1$ $-1$)$/\sqrt{12}$\,,
is the dominant component of atomic displacements
through out the entire speculating range
in BaTiO$_3$, as it is well known.
We find that it is useful
to analyze normalized atomic displacements of PbTiO$_3$
with new three components,
the $A$-O$_1$,O$_2$ closing mode
$\bm{\xi}_{A-{\rm O}_1,{\rm O}_2} = {}^t$(2 0 $-1$ $-1$ 0)$/\sqrt{6}$\,,
the $B$-O$_3$ closing mode
$\bm{\xi}_{B-{\rm O}_3} = {}^t$(0 1 0 0 $-1$)$/\sqrt{2}$\,,
and the rest mode
$\bm{\xi}_{\rm rest} = {}^t$($-2$ 3 $-2$ $-2$ 3)$/\sqrt{30}$\,.
The dominant $A$-O$_1$,O$_2$ closing mode
has its maximum at around $u_z=0.6087$,
at which the total energy becomes minimum,
as shown in Fig~\ref{fig:PbTiO3Mode}.
This can be understood as follows: from $u_z=0$ to $u_z=0.6087$
the covalent bonding between A-O$_1$,O$_2$ becomes stronger,
but beyond  $u_z=0.6087$, rigid-sphere-like ionic repulsion between
them becomes stronger.
It is consistent with Cohen's description~\cite{Cohen:Nature:1992}
and Kuroiwa {\it et al.}'s experimental work~\cite{Kuroiwa:X-ray:Pb:O:Covalency:2001}.

In summary,
we contrived an {\it ab initio} structure optimization technique
to determine the valley line on a total-energy surface for
zone-center distortions of
ferroelectric perovskite oxides
and applied it to
BaTiO$_{3}$ and
PbTiO$_{3}$.
(a) Our new optimization technique
picks up the tetragonal equilibrium structure
and gives its correct energy gain in principle.
(b) The results of numerical calculations show that
total energy can be expressed as a fourth-order function
of the amplitude of atomic displacements $u_z$
in BaTiO$_3$ but not in PbTiO$_3$.
(c) Our new optimization technique can {\em automatically}
evaluate the total energy as a function of $u_z$
and coupling between atomic displacements and strains, i.e.,
the quadratic or nearly quadratic $u_\alpha$-dependences of strains.
This is an advantage of our technique
compared with King-Smith and Vanderbilt's scheme.

Computational resources
were provided by the Center for Computational Materials Science,
Institute for Materials Research, Tohoku University.

%%%%%%%%%%%%%%%%%%%%%%%%%%%%%%%%%%%%%%%%%%%%%%%%%%%%%%%%%%%%%%%%%%%%%%%%%%%%%%%%

\end{document}